\documentclass[12pt]{article}

\usepackage{amsmath,amssymb,graphicx,cite} 
\usepackage{epsf}

\newcommand{\arline}{\nonumber \\}

\newcommand{\pd}{\partial}

\newcommand{\Tr}{\mathrm{Tr~}}

\renewcommand{\theequation}{\thesection.\arabic{equation}}

\begin{document}
\begin{titlepage}
\begin{flushright}
{\tt FERMILAB-PUB-07-136-T} \\
{\tt CLNS 06/1966}
\end{flushright}

\vskip.5cm
\begin{center}
{\huge \bf Radion Phenomenology in Realistic Warped Space Models}
\vskip.2cm
\end{center}

\begin{center}
{\bf {Csaba Cs\'aki}$^a$, {Jay Hubisz}$^b$, and {Seung J. Lee}$^a$} \\
\end{center}
\vskip 8pt

\begin{center}
$^{a}$ {\it Institute for High Energy Phenomenology\\
Newman Laboratory of Elementary Particle Physics\\
Cornell University, Ithaca, NY 14853, USA } \\

\vspace*{0.3cm}

$^b$ {\it Fermi National Accelerator Laboratory\\
P.O. Box 500, Batavia, IL 60510, USA}

{\tt  csaki@lepp.cornell.edu, hubisz@fnal.gov, sjl18@cornell.edu}
\end{center}

\vglue 0.3truecm

\begin{abstract}
\vskip 3pt \noindent

We investigate the phenomenology of the Randall-Sundrum radion in
realistic models of electroweak symmetry breaking with bulk gauge
and fermion fields, since the radion may turn out to be the lightest
particle in such models. We calculate the coupling of the radion in such
scenarios to bulk fermion and gauge modes. Special attention needs
to be devoted to the coupling to massless gauge fields (photon,
gluon), since it is well known that
loop effects may be important for these fields. We also present a detailed
explanation of these couplings from the CFT interpretation. We then
use these couplings to determine the radion branching fractions and
discuss some of the discovery potential of the LHC for the radion.
We find that the $\gamma\gamma$ signal is enhanced over most of the
range of the radion mass over the $\gamma\gamma$ signal of a SM
Higgs, as long as the RS scale is sufficiently low. However, the
signal significance depends strongly on free parameters that characterize
the magnitude of bare brane-localized kinetic terms for the massless
gauge fields. In the absence of such terms, the signal can be
be enhanced over the traditional RS1
models (where all standard model fields are localized on the IR
brane), but the signal can
also be reduced compared to RS1 if the brane localized terms are
sizeable. We also show that for larger radion masses, where the
$\gamma\gamma$ signal is no longer significant, one can use the usual 4
lepton signal to discover the radion.
\end{abstract}

\end{titlepage}

\newpage


\section{Introduction}
\label{intro}
\setcounter{equation}{0}
\setcounter{footnote}{0}


There has been a lot of attention devoted to models of physics above
the weak scale utilizing warped extra dimensions. The first
proposal~\cite{RS1} of Randall and Sundrum (RS1) solves the
hierarchy problem by localizing all standard model (SM) particles on
the IR brane. 
Much research has been done on understanding possible mechanisms for
radius stabilization and the phenomenology of the radion field in
this model~\cite{GW,CGRT,GRW,TanakaMontes,CGK,Gunion,Korean}. The
motivation for studying the radion is twofold.  First, the radion
may turn out to be the lightest new particle in the RS-type setup,
possibly accessible at the LHC.  In addition, the phenomenological
similarity and potential mixing of the radion and the Higgs boson
warrant detailed study in order to facilitate distinction between
radion and Higgs signals at colliders.

By now it is quite clear that, in order to make a realistic model of
electroweak symmetry breaking (EWSB) in an RS-type scenario, the
model needs to be modified by extending the gauge fields and
fermions into the bulk. This is forced by constraints from flavor
physics and electroweak precision tests of the SM. Since the cutoff
scale on the IR brane is of the order of a few TeV, one would expect
dimension 6 operators localized on the TeV brane to be suppressed by
this relatively low scale, giving unacceptably large corrections to
electroweak precision observables such as the T-parameter,
and also to flavor changing neutral currents. Another way
to see this is to invoke the AdS/CFT
correspondence~\cite{CFT,CFTgauge,CFTfermions}. The conformal field
theory (CFT) interpretation of the 2-brane RS1 model is a 4D CFT
that is spontaneously broken at the TeV scale. At the scale of
conformal symmetry breaking, the broken CFT produces a SM where all
of the fields are composites, and since the scale of compositeness
is low one would again expect large corrections to precision
observables.

In order to overcome this problem first the gauge fields were moved
to the bulk~\cite{bulkgauge}, but $S$ and $T$ still remained
large~\cite{RSeff}. A more realistic
approach can be found by moving both gauge fields and fermions into
the bulk, and incorporating a custodial
symmetry~\cite{ADMS}. Variations of this
basic setup include Higgsless models (with boundary condition
EWSB)~\cite{higgsless}
and holographic composite Higgs models~\cite{holoHiggs}.

While the gauge and matter sector model building has advanced
tremendously, little attention has been paid to the modification of
the radion
physics due to the change in the model structure (with the exception
of~\cite{Rizzo}).
The aim of this paper
is to lay down the groundwork for finding the basic properties of
the radion in one of these realistic warped EWSB models. In Sec. 2
we review the radion mode and the general expression for its
tree level coupling. In Sec. 3 we find the couplings to bulk
fermions, including those localized on the UV and IR branes. In Sec.
4 we investigate the couplings to gauge bosons. Particular attention
needs to be paid to massless gauge bosons, since some of the
calculable one-loop contributions turn out to be important. This has
to be investigated carefully, since important
production and discovery channels at the LHC include couplings to
massless gauge fields. In Sec.~5 we review the CFT interpretation of
the radion and use this to re-derive the various couplings to
massive and massless fields. In Sec.~6 we present the branching
fractions of the radion as a function of the parameters of the 5D
model, while in Sec.~7 we show some of the possible discovery
channels at the LHC. We conclude in Sec.~8, and present Appendices
clarifying the role of boundary kinetic terms for fermions, the CFT
interpretation of fermion masses for bulk fields, and an explicit 5D
loop calculation illustrating that renormalization of the brane
induced gauge kinetic terms captures the leading one loop effects of the
radion coupling to massless gauge bosons.


\section{The radion solution}
\setcounter{equation}{0}
\setcounter{footnote}{0}

Throughout this paper, we use the conventions
of~\cite{CGK,CGHST}. The $z$ coordinate will always refer to the conformally
flat AdS background with $R<z<R'$, and the AdS curvature is $R=1/k$, while the
$y$ coordinates are given by $y=R \log z/R$. The AdS metric including the scalar perturbation
$F$ corresponding to the effect of the radion is given in these coordinate systems by
 \begin{equation}
ds^2 = e^{-2 (A + F)} \eta_{\mu\nu} - (1+2F)^2 dy^2 = \left(\frac{R}{z}\right)^2\left( e^{-2F} \eta_{\mu\nu} dx^\mu dx^\nu  - (1+2F)^2 dz^2
\right),
\end{equation}
where $A(y) =k\,y$. Note that the perturbed metric is no longer
conformally flat, even in $z$ coordinates.
At linear order in the flutuation, $F$, the
metric perturbation is given by
\begin{equation}
\delta( ds^2 ) \approx -2 F \left( e^{-2A}
\eta_{\mu\nu} dx^\mu dx^\nu + 2~dy^2 \right) = -2 F \left( \frac{R}{z}
\right)^2 \left( \eta_{\mu\nu}dx^\mu dx^\nu + 2 dz^2\right).
\label{radionpert}
\end{equation}

In the absence of a stabilizing mechanism, the radion is precisely
massless, however it was shown that the addition of a bulk scalar
field with a vacuum expectation value (vev) leads to an effective
potential for the radion after taking into account the back-reaction
of the geometry due to the scalar field vev profile~\cite{GW}.  In
our analysis that follows, we assume that this backreaction is small,
and does not have a large effect on the 5D profile of the radion.

The radion is assumed to take the form $F(x,z) = r(x) {\rm R}(z)$,
where the form of $R(z)$ is determined by imposing that the metric
solve Einstein's equations, and that $r(x)$ be a canonically
normalized 4D scalar field after integrating out the extra dimension.

In the limit of small back-reaction, the relation between the
canonically normalized 4D radion field $r(x)$ and the metric
perturbation $F(z,x)$ is given by
\begin{equation}
F(z,x) = \frac{1}{\sqrt{6}} \frac{R^2}{R'} \left( \frac{z}{R}
\right)^2 r(x) =
\frac{r(x)}{\Lambda_r}\,\left(\frac{z}{R'}\right)^2,
\end{equation}
with
\begin{equation}
\Lambda_r \equiv \frac{\sqrt{6}}{R'}.
\end{equation}

In order to find the interaction terms between the radion and the
SM fields to linear order in the radion, we can use the
fact that the
energy momentum tensor of the matter fields is just the linear variation
of the action with respect to the metric, thus
\begin{equation}
S_{\mathrm{radion}} = -\frac{1}{2} \int d^5x \sqrt{g} T^{MN} \delta
g_{MN}.
\end{equation}
Plugging in for the expression for $\delta g_{MN}$ corresponding to
the radion fluctuation from Eq. (\ref{radionpert}), we find the
interactions of the radion are given by
\begin{equation}
\label{eq:5DTrace} S_{\mathrm{radion}} = \int d^5x\sqrt{g} \left[ F
\left( \Tr T^{MN} - 3 T^{55} g_{55}
  \right) \right].
\end{equation}
Note that the radion couples to a 5D scale invariant object, which can be
thought as an effective 4D trace when integrated over the
z-direction. If all the SM fields are localized on the IR brane,
these become the radion interactions of the RS1 model,
$\frac{r}{\Lambda_r}\,T^{\mu}_{\,\mu}$, which is proportional to the
4D trace of the brane localized energy-momentum tensor~\cite{GW,CGRT,GRW,TanakaMontes,CGK,Gunion,Korean}.

\section{Radion Couplings to Fermions}
\setcounter{equation}{0}
\setcounter{footnote}{0}

The 5D action for bulk fermions can be written as\footnote{For general discussions
of fermions in 5D warped space see~\cite{GN,GP,HuberShafi,CGHST}.}
\begin{equation}
\label{eq:BulkAction} S = \int d^5 x \sqrt{g} \left( \frac{i}{2} (
\bar{\Psi}\, \Gamma^M D_M \Psi - D_M \bar{\Psi}\, \Gamma^M  \Psi ) +
m \bar{\Psi} \Psi \right),
\end{equation}
where $\Psi$ is the 5D Dirac spinor, and M=0,1,2,3,4,5. Here the
$\Gamma$ matrices are  
given by $\Gamma^M=\gamma^{a} e^M_a$, where the $\gamma^a$ are the
ordinary 
$\gamma$-matrices with $\gamma^5= i {\rm diag} (1_2,-1_2)$, and the 5D
vielbein 
defined by $e^M_a e^N_b \eta^{ab}=g^{MN}$ is given for the metric
including the radion 
fluctuation (to linear order in $F$) by
\begin{equation}
e^M_a ={\rm diag} \frac{z}{R} (1+F,1+F,1+F,1+F,1-2 F).
\end{equation}
The covariant derivative is given by $D_M=\partial_M+\frac{1}{2}
\omega_{bcM}\sigma^{bc}$ 
where the $\omega$'s are the spin connections. This action can be also 
written as 
 \begin{equation}
\label{eq:BulkAction2} S = \int d^5 x \sqrt{g} \left(
e^M_a\frac{i}{2} ( \bar{\Psi}\, \gamma^a \partial_M \Psi -
\partial_M \bar{\Psi}\, \gamma^a  \Psi +\omega_{bcM}\bar{\Psi}
\frac{1}{2} \{\gamma^a,\sigma^{bc}\} \Psi) + m \bar{\Psi} \Psi
\right).
\end{equation}
The advantage of writing the action in this form is that one can show
that the contribution 
involving the spin connections will be vanishing for diagonal metrics
(and vielbeins), as 
in our case.

In order to find the interaction of fermions with the radion we expand
this action to linear order in $F$. We first separate the  
bulk Dirac fermion into two component spinors as
\begin{equation}
\Psi = \left(
\begin{array}{c}
\chi_{\alpha}  \\
\bar{\psi}^{\dot{\alpha}}
\end{array}
\right),
\end{equation}
where $\chi$ is the left handed spinor and $\bar{\psi}$ is the right handed. For their
interactions with the radion we find (introducing the usual notation $c=mR$):
\begin{equation}
\frac{r}{\sqrt{6}}\frac{R^2}{R'^2} \int dz
\left(\frac{R}{z}\right)^2 \left[ -i (\psi \sigma^\mu \partial_\mu
\bar{\psi} +\bar{\chi}\bar{\sigma}^\mu\partial_\mu \chi)+ 2 (\psi
\overleftrightarrow{\partial_5}  \chi -\bar{\chi}
\overleftrightarrow{\partial_5} \bar{\psi}) +\frac{2 c}{z} (\psi
\chi+\bar{\chi} \bar{\psi})\right]. \label{radioncoupling1}
\end{equation}
We are interested in the interaction terms between the 4D Kaluza-Klein
(KK) modes of the
fermions with the radion. The KK decomposition is, as usual, given by
\begin{eqnarray}
        \label{eq:DiracKK}
\chi_  = \sum_n g^n (y)\, \chi^n (x), \\
\bar{\psi} = \sum_n f^n_i (y)\, \bar{\psi}^n (x).
\end{eqnarray}
where $\chi^n (x)$ and $\bar{\psi}^n (x)$ are four dimensional spinors satisfying the 4D Dirac equation
\begin{eqnarray}
-i \bar{\sigma}^{\mu} \partial_\mu \chi_{n} + m_n\, \bar{\psi}_n = 0, \\
-i \sigma^{\mu} \partial_\mu \bar{\psi}_n + m_n\, \chi_{n} = 0.
\end{eqnarray}
Using this, Eq.~(\ref{radioncoupling1}) can be rewritten as
\begin{equation}
\frac{r}{\sqrt{6}} (\psi_n \chi_n +\bar{\chi}_n\bar{\psi}_n)
\int dz \left(\frac{R}{z}\right)^2 \frac{R^2}{R'} \left[ -\frac{m_n}{2}
  \left( f_n^2 + g_n^2\right)+ 2 (f_n g_n'-f_n'g_n+ \frac{c}{z} f_ng_n)
\right],
\label{radionferm}
\end{equation}
where we have assumed that the two fermions coupling to the radion are
at the same level in the KK tower.  It is easy to extend this formula
to calculate the coupling of the radion to two different KK mode
fermions, but for the purposes of this paper we are most interested
in the couplings of the radion to the SM fermions (the lightest
elements of the KK tower).  In this case, Eq.~(\ref{radionferm}) is sufficient.

One subtlety that one needs to clarify is the effect of adding
brane localized interactions for the fermions.  First, boundary mixing of
bulk fermions is what normally leads to masses for the SM fermions, so in
general, Eq.~(\ref{radionferm}) must be summed over all bulk fermions
that mix to form the mass eigenstates.  In addition,
it appears
naively that localized mass terms will also give a direct
contribution to the coupling. However, in Appendix A we show that a
careful treatment of the boundary conditions implies that the
contributions from the localized terms actually cancel against the
induced wave function discontinuity contributions to the stress-energy
tensor at the TeV brane. Thus
Eq.~(\ref{radionferm}) is the full expression for the fermion-radion
coupling.

\subsection{Approximate radion couplings to fermions in a simplified model of fermion masses}

In order to evaluate the expression (\ref{radionferm}) we
need to understand the basic properties of the mechanism responsible
for generating the fermion masses. We consider the radion coupling to
fermions in the simple example
of a
bulk SU(2)$_L\times$U(1)$_Y$ gauge group with the usual SM quantum
number assignments for the bulk Dirac fermions. The boundary
conditions will then be chosen such that from the SU(2)$_L$ doublets
the left handed ($\chi$) fermions have zero modes, while from the
singlets it is the right handed ($\psi$) fermions that have zero
modes. If we denote the SU(2) doublet fields as L-fields and the
singlet fields as R-fields\footnote{L and R correspond only to
chiralities of the zero modes; the L and R bulk fields contain both left handed and right handed
fermions in the KK expansion}, then the BC's providing the proper zero modes are
$\psi_L|_{z=R,R'}=\chi_R|_{z=R,R'}=0$. In this simple model, the electroweak symmetry is
broken via the Higgs, which is localized on the TeV brane, and which
generates masses for the zero modes. The localized Higgs will
generate a TeV-brane localized Dirac mass: $M_D R' \bar{\Psi}_L
\Psi_R$. In this model, for every fermion there are bulk mass parameters
$c_L$ for the doublets and a separate $c_R$ for every right handed
field which characterize the profiles of the zero modes. With the
conventions used in this paper, the wave functions of the fermions
are localized on the Planck brane for $c_L>1/2$ (and for
$c_R<-1/2$), while in the opposite case they are localized on the
TeV brane. In addition there is a separate Dirac mass term $M_D$ for
every fermion.

The radion coupling to SM fermions should be proportional to the
physical mass of this field.  Generically, these lowest lying modes
are considerably lighter than the scale $1/R'$, making it useful to find the
radion-fermion coupling as an expansion in $m R'$.  The expression
for the lightest eigenvalue is found to be:
\begin{equation}
m^2 = M_D^2 \frac{ (1+2c_R) (1-2c_L) }{\left( 1- \lambda^{1+2c_R}
\right) \left( 1 - \lambda^{1-2c_L} \right) },
\label{eq:fermmass}
\end{equation}
where we have defined $\lambda \equiv R/R'$. The light fermions are
usually assumed to be localized around the Planck brane ($c_L>1/2,
c_R<-1/2$). For this case, the coupling in Eq.~(\ref{radionferm})
simplifies to
\begin{equation}
 \frac{r}{\Lambda_r} m (c_L-c_R)  \label{lightfermcoupl}
\end{equation}
after summing over the bulk L and R fields.

Note that, for the light fermions, the interactions with the radion
depend on the bulk profiles. In order for the top to be
sufficiently heavy we need to assume that its wave function is
localized on the TeV brane, that is $c_L<1/2$ and $c_R>-1/2$.   As
expected for this TeV brane localized case, the leading coupling is
equal to the usual result from the brane localized mass terms in the
RS1 model, and is insensitive to the bulk mass parameters:
\begin{equation}
 \frac{r}{\Lambda_r} m. \label{topquarkcoupl}
\end{equation}

\subsection{Radion coupling to fermions in models with custodial SU(2)}

Realistic models of EWSB with bulk fermions need to incorporate a
custodial SU(2) symmetry in order to protect the $T$-parameter from
large corrections. This is achieved by gauging
SU(2)$_L\times$SU(2)$_R\times$U(1)$_{B-L}$ in the bulk of AdS$_5$,
and then breaking SU(2)$_L\times$SU(2)$_R\to$SU(2)$_D$ on the TeV
brane either via a localized Higgs or boundary conditions, while on
the Planck brane SU(2)$_R\times$U(1)$_{B-L}\to$U(1)$_Y$. This setup
has been first suggested in~\cite{ADMS} for RS-type models, and
applied to Higgsless models in~\cite{higgsless}. This implies a
modification of the model of fermion masses discussed in the
previous section. The simplest possibility is to put the two right
handed bulk fields into a doublet of SU(2)$_R$. Similar to the
previous case without custodial SU(2), the zero modes obtained from
the orbifold projections will pick up a mass once a Dirac mass
term $M_D \bar{\Psi}_L \Psi_R$ is added on the TeV brane. In order
to split the up and down type quarks (and the charged leptons from
the neutrinos) one can either introduce a brane kinetic term (BKT) for the
R-fields on the Planck brane or, for the neutrinos, a Majorana mass
for the right handed component.  The parameters of this model are:
$c_L,c_R,M_D$ and a parameter $\alpha$ characterizing the size of the
Planck brane induced kinetic term for the SU(2)$_R$ fields (or $M_M$
for a RH neutrino Majorana mass).

Another possible model for the fermion masses with custodial SU(2)
is to introduce a separate SU(2)$_R$ doublet for every SM field. In
this case the correct chiral spectrum is obtained by assigning the
following parities:
\begin{itemize}
\item $(+,+)$ for $\chi_L, \psi_{u_R}$ and $\tilde{\psi}_{d_R}$
\item $(-,-)$ for $\psi_L, \chi_{u_R}$ and $\tilde{\chi}_{d_R}$
\item $(+,-)$ for $\chi_{d_R}$ and $\tilde{\chi}_{u_R}$
\item $(-,+)$ for $\psi_{d_R}$ and $\tilde{\psi}_{u_R}$
\end{itemize}
In this case one can add a separate Dirac mass on the TeV brane
mixing the left handed doublet with either of the right handed ones.
A mixing term between the right handed fields (which would be
allowed by the gauge quantum numbers) is vanishing due to the parity
assignments. In this case the parameters of the model are one $c_L$
and a separate $c_R$ for every SU(2)$_R$ field, and a separate Dirac
mass on the TeV brane. Thus the parameters in this model actually
match the parameters in the model without custodial SU(2) symmetry.

We can again evaluate the approximate expression for the
fermion-radion coupling for both of these models. For the case of
the model with the three multiplets the general expression for the
coupling is given by
\begin{equation}
\frac{r}{\Lambda_r} m \frac{c_R-c_L+\frac{1}{2} (1-2 c_R) \lambda^{2
c_L-1}-\lambda^{2c_L-2c_R-2}}{(1-\lambda^{2c_L-1})(1-\lambda^{-1-2
c_R})}.
\end{equation}
For the case relevant for the light fermions we get the same
approximate expression (\ref{lightfermcoupl}) as in the simple model
and similarly for the case relevant to the top quark we find
(\ref{topquarkcoupl}). However, for the bottom quark (assuming
$c_L<1/2, -0.68<c_R<-1/2$) we find a coupling given by
\begin{equation}
\frac{r}{\Lambda_r} \frac{m}{2}(2c_R-1).
\end{equation}
One can check that, to leading order in the masses, the same
approximate expressions apply in the model with just two bulk
fermions where there are Planck brane localized kinetic terms for the
SU(2)$_R$ fields.

\section{Radion couplings to gauge bosons}
\setcounter{equation}{0}
\setcounter{footnote}{0}

We expect to find the most important effects in this sector. The
reason is that once the gauge field is a(n approximate) bulk zero mode
its wave function is not peaked on the TeV brane, so there is no
limit in which this setup reproduces the naive RS1 model. This will
result in the main new effect: the existence of tree level
couplings of the radion to the bulk kinetic terms of the gauge bosons.
For the massive gauge bosons, the coupling proportional to
the mass is dominant over the new kinetic term coupling.
For massless gauge bosons, however, the tree level coupling to the field
strength squared is often the dominant effect, and one has to add the
effects from the bulk and brane localized kinetic terms.

\subsection{Radion couplings to massive gauge bosons}

First we consider the simpler case of massive gauge bosons. Just
as in the case of fermion couplings, we treat the Higgs as
completely localized on the IR brane. The action for the massive
gauge bosons is then given by
\begin{equation}
S_{\mathrm{W}} = \int \sqrt{g} \Big(-\frac{1}{2}
g^{MN}g^{KL}W^{\dagger}_{MK}W_{NL}+ g^{\mu\nu}\frac{\delta(
z-R')}{\sqrt{g_{55}}}\, \left(g_5\,
v\right)^2W_\mu^{\dagger}W_\nu\Big),
\end{equation}
and
\begin{equation}
S_{\mathrm{Z}} = \int \sqrt{g} \Big(-\frac{1}{4}\,
g^{MN}g^{KL}Z_{MK}Z_{NL}+ \frac{1}{2}\,g^{\mu\nu} \frac{\delta(
z-R')}{\sqrt{g_{55}}}\,( \sqrt{g_{5}^{2}+{g'_{5}}^2} \,v)^2 Z_\mu
Z_\nu\Big),
\end{equation}
where $v$ is the localized Higgs vev, which can be different from
the SM vev in two ways: first, one needs to rescale by the usual RS
warp factor in AdS space, second, in principle a large fraction of
the gauge boson mass could arise from the bulk curvature of the
gauge boson wave functions as in Higgsless models.  In
either case, the dominant couplings to the radion are given by:
\begin{equation}
{\cal{L}} = -\frac{r}{\Lambda_r}\,\left(2M^2_W\,W^{+}_{\mu}\,W^{\mu
-}+M_Z^2 Z_{\mu}\,Z^{\mu}\right). \label{WZcoupl}
\end{equation}

If the $W,Z$ masses come from a
localized Higgs vev, the corrections to the coupling in Eq.~(\ref{WZcoupl})
are then given by 
\begin{equation}
-\frac{r}{4\Lambda_r\log{\frac{R'}{R}\,}} (2W^{\dagger}_{\mu\nu}W^{\mu\nu}+Z_{\mu\nu}Z^{\mu\nu})
+\frac{M_W^4\,R'^2}{\Lambda_r}\,\log{\frac{R'}{R}\,}
r\,W_\mu\,W^{\mu}+\frac{M_Z^4\,R'^2}{2\Lambda_r}\,\log{\frac{R'}{R}\,}
r\,Z_\mu\,Z^{\mu},
\label{corrmassg}
\end{equation}
where we have used the wave function for the $W$ boson given in~\cite{RSeff}:
\begin{equation}
W^\mu(z,x)\simeq\frac{1}{\sqrt{R\log{\frac{R'}{R}\,}}}\,
\Big(1+\frac{M_W^2}{4}\,(z^2-R'^2-2z^2\log{\frac{R'}{R}\,}+2R'^2\log{\frac{R'}{R}\,})\Big)
W^\mu(x).
\end{equation}
A similar formula applies for the $Z$ boson.  We have assumed
that there are no BKTs for the $W$ and $Z$.  Note that these
corrections are dependent on the specifics of EWSB, and will be
different, for example, in Higgsless models of EWSB.  The
coupling to the field strengths in the first term of
Eq.~(\ref{corrmassg}) is a new feature of bulk RS gauge fields that
plays a significant role only for momentum transfer well above the
electroweak mass scale. 

\subsection{Radion couplings to the massless gauge bosons}

For the massless gauge bosons, there is no large coupling to the
radion as there are no brane localized mass terms. In order to take
potentially large loop effects into account we also consider brane
localized kinetic terms for the gauge fields. The action for the
massless gauge boson is then given by (note that, for this discussion,
is it convenient to use non-canonically normalized gauge fields)
\begin{equation}
S_{\mathrm{massless}} = -\frac{1}{4g_5^2}\int d^5 x \sqrt{g} F_{MN}F^{MN}
-\frac{1}{4} \int d^4 x \sqrt{g_{UV,IR}} \tau_{UV,IR} F_{\mu\nu}F^{\mu\nu}
\end{equation}
where $F$ is either the photon or the gluon, and
$\tau_{UV},\tau_{IR}$ parameterize the Planck and TeV brane induced
kinetic terms. With this definition, the tree-level matching relation
between the 4D and 5D couplings is given
by~\cite{RandallSchwartz,GR,Choi,ADS,Italianrunning}
\begin{equation}
\frac{1}{g_4^2}=\frac{R\log \frac{R'}{R}}{g_5^2} +\tau_{UV}+\tau_{IR}
\end{equation}
Plugging in the gauge boson and radion wave functions, we find
that the bulk contribution to the coupling is
\begin{equation}
-\frac{R}{R \log{\frac{R'}{R}\ +(\tau_{UV}+\tau_{IR})g_5^2}}
\frac{r}{4\Lambda_r}
F_{\mu\nu} F^{\mu\nu} =-\frac{R g_4^2 }{g_5^2}
\frac{r}{4\Lambda_r}
F_{\mu\nu} F^{\mu\nu}.
\label{masslessGBcoupl}
\end{equation}
Note that the above expression gets a direct contribution only from
the bulk, since the induced 4D kinetic terms are scale invariant at
tree level. This formula already incorporates some of the one-loop
corrections as well: it contains the loop effects corresponding to
the renormalization of the local 5D operators (bulk + brane
localized kinetic terms). In particular, potentially large
linearly divergent contributions to the radion coupling from one-loop
bulk gauge coupling renormalization and bulk triangle diagrams 
vanish (as we show in Appendix C) after assuming appropriate bulk
counterterms and using the 
renormalized values for the bulk and brane localized parameters. In
addition there can be 
brane localized direct contributions to the radion-gauge boson
coupling which we will discuss, and possible (subleading) non-local
effects which we neglect.

We now add the effect of the localized trace anomalies and of loop
induced couplings due to the zero modes.  There are no large loop
induced bulk couplings.  The reason for this is that
in the bulk there is a tree level coupling between the radion
and the gauge fields.  Therefore loop effects merely
renormalize this tree level operator.  On the branes, there is no
allowed tree level coupling, thus the loop effects are finite and
have to be included separately.  Since the wave function of the
radion is negligible on the UV brane, we only add the effects on the
IR brane. The localized trace anomaly on the IR brane is exactly the
same effect as the entire radion coupling to massless gauge bosons
in the original RS1 model, as studied in~\cite{GRW}, except we need
to add it only for 
the fields that are localized on the IR brane.  Thus the coupling
due to the localized trace anomaly will be
\begin{equation}
-\frac{r}{\Lambda_r} \frac{b_{IR}\alpha}{8\pi} F_{\mu\nu}F^{\mu\nu}.
\end{equation}
where $b_{IR}$ is the $\beta$-function coefficient of the light fields localized on
the IR brane. In our case
these are the top quark pair $\left( t_{L},\, t_{R}\right)$, left-handed bottom quark $\left(b_{L}\right)$ and the Higgs. Thus for
the gluon $b^3_{IR}=b^{3,t_L}+b^{3,t_R}+b^{3,b_L}=-1$, while for the
photon $b^{EM}_{IR}=b^{EM,H}+N_c b^{EM,t_L}+N_c b^{EM,t_R}+N_c b^{EM,b_L}=-7/3$.

\begin{figure}
\begin{tabular}{ccccc}
   \includegraphics[width=.17\hsize]{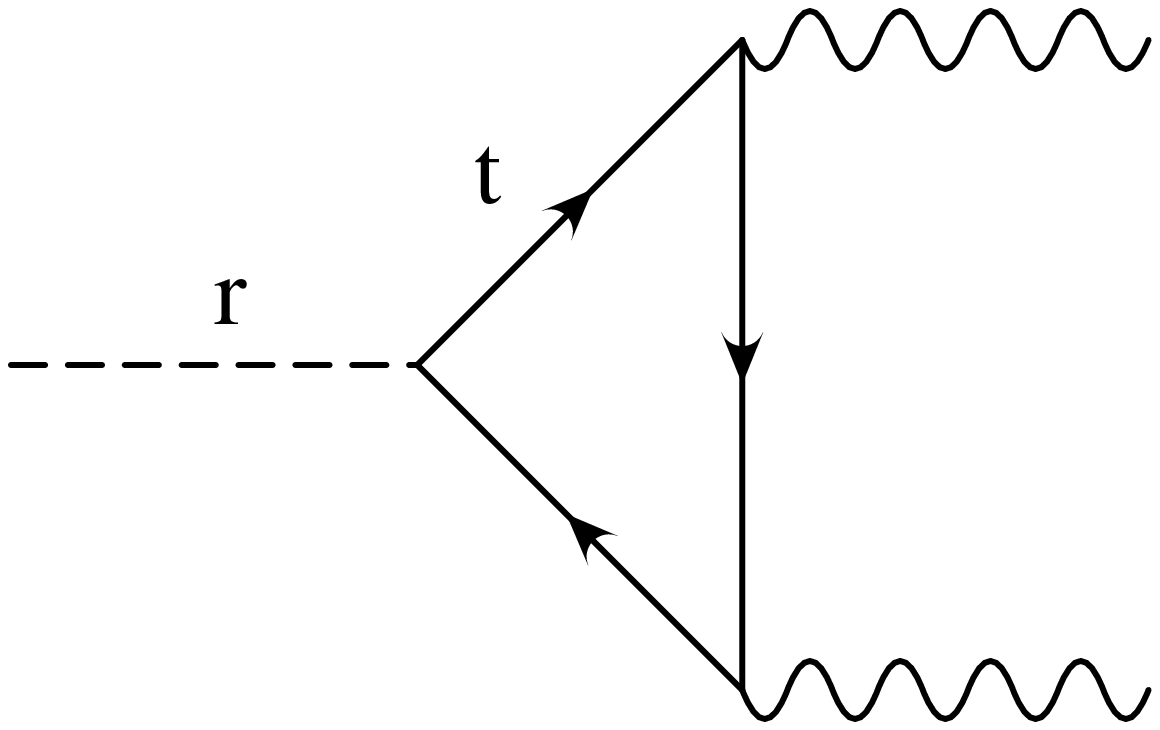} & \includegraphics[width=.17\hsize]{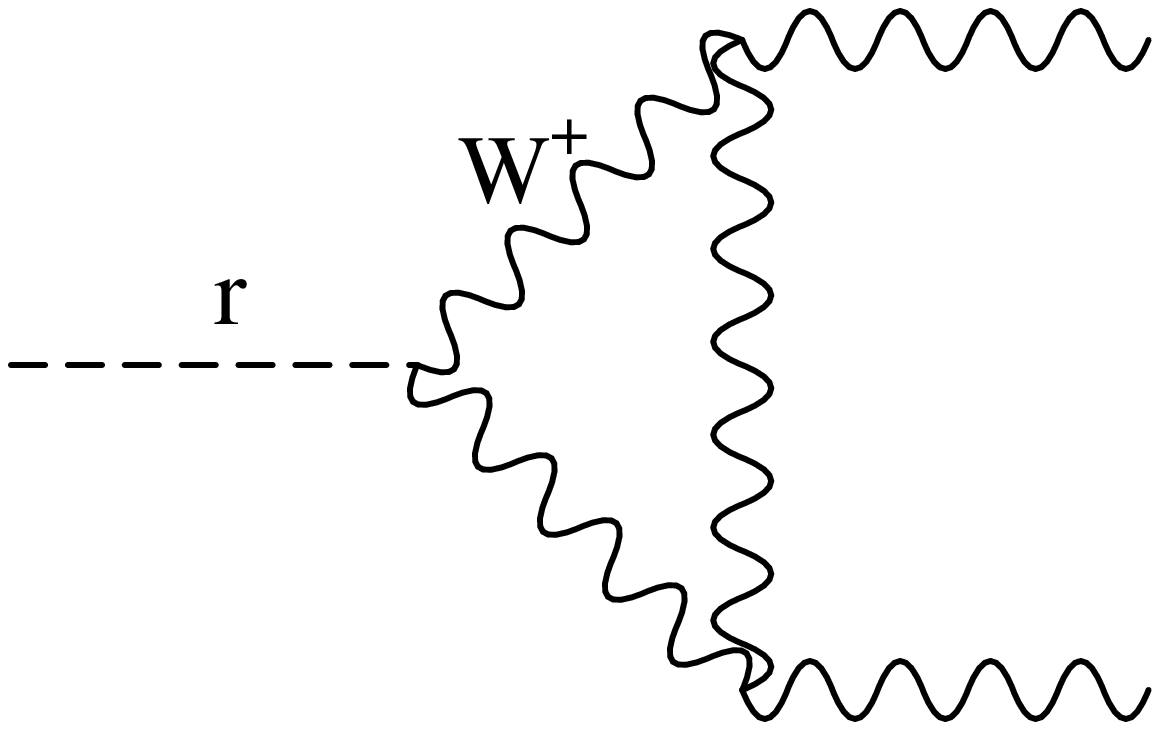} & \includegraphics[width=.17\hsize]{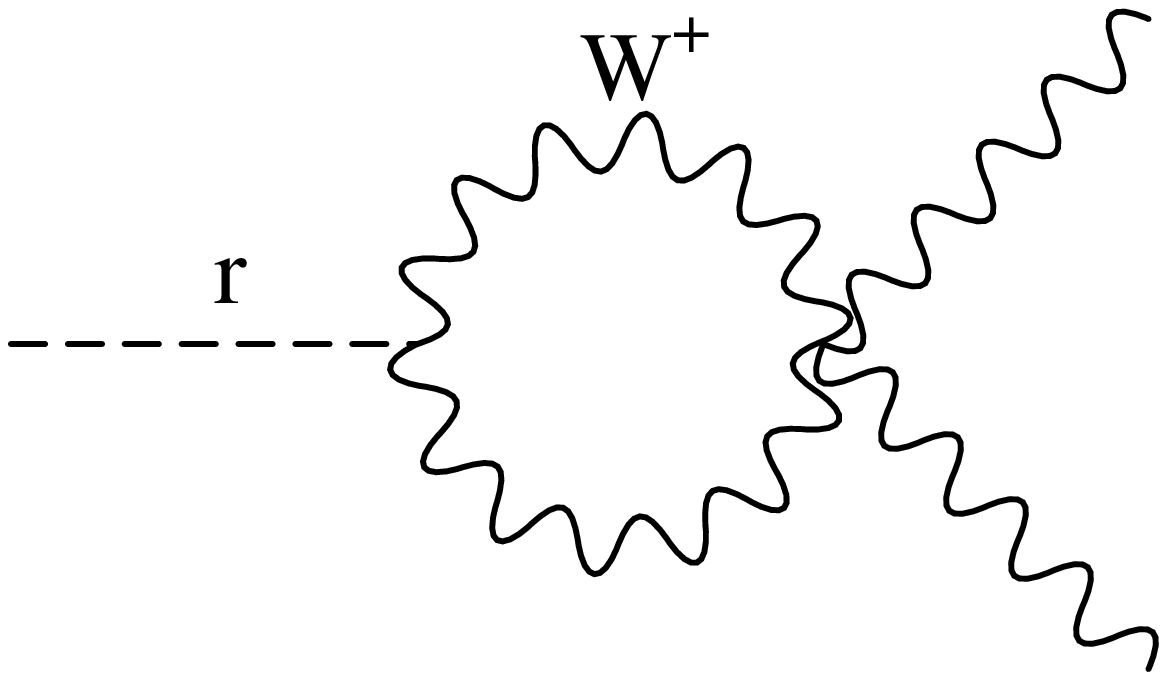} & \includegraphics[width=.17\hsize]{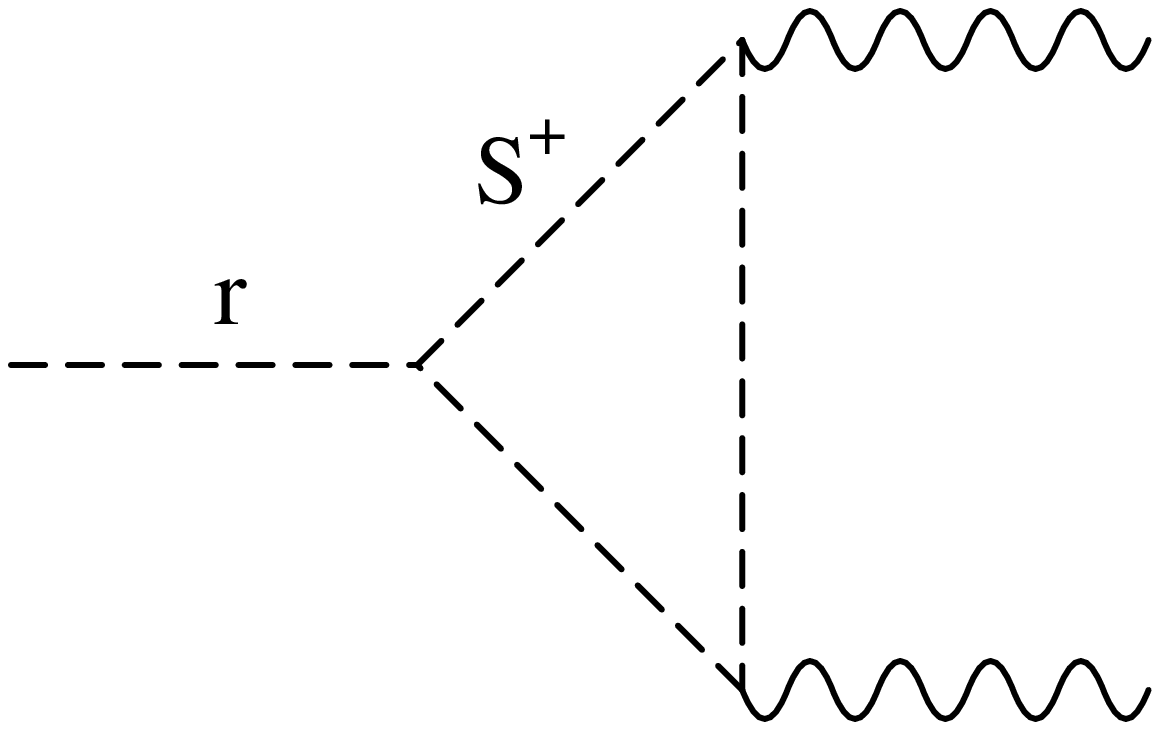} & \includegraphics[width=.17\hsize]{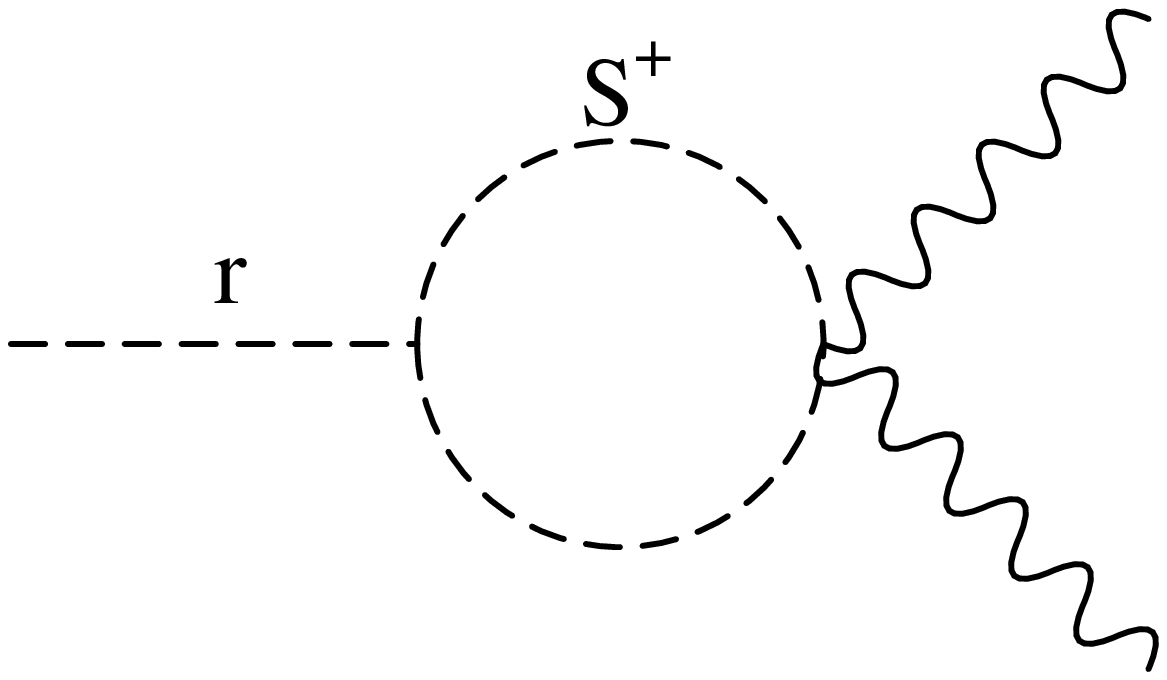} \\

\end{tabular}
\caption{Triangle diagrams \label{triangle}}
\end{figure}

In addition, there are triangle diagrams, as shown in
Fig\ref{triangle}, involving either the gauge fields, the Higgs or
the top quark that introduce a finite contribution to the coupling.
We calculate this in the low energy effective 4D theory.  This is
again the analog of the effect of the straight RS1 contribution as
explained in~\cite{GRW}.  Noting that the leading contributions to the
couplings to heavier fields, such as $W$ and top, are the same as in the
RS1 model, 
the loop induced brane localized coupling due to the top quark is given by
\begin{equation}
 \mu_t F_{1/2}(\tau_t)
\frac{\alpha}{8\pi}\frac{r}{\Lambda_r}F_{\mu\nu}F^{\mu\nu}
\end{equation}
where $\mu_t =1/2$ for the gluon and $Q_t^2$ for the photon. In this
expression $\tau_t=4 (m_t/m_r)^2$, and $F_{1/2}(\tau )=-2\tau
[1+(1-\tau )f(\tau )]$, with $f(\tau ) = [\sin^{-1} (1/\sqrt{\tau
})]^2$ for $\tau >1$. The important property of $F_{1/2}(\tau )$ is
that, for $\tau >1$, it very quickly saturates to $-4/3$, and to $0$
for $\tau <1$. Similarly
the massive $W$  will contribute to the photon coupling
\begin{equation}
F_{1}(\tau_W)\frac{\alpha}{8\pi}\frac{r}{\Lambda_r}F_{\mu\nu}F^{\mu\nu},
\end{equation}
where $\tau_W=4 (m_W/m_r)^2$, $F_1(\tau )=2+3\tau +3\tau (2-\tau )f(\tau ) $ and $F_1(\tau )$
saturates quickly to $7$ for $\tau >1$, and to $0$ for $\tau < 1$.

So the combined induced coupling for the gluon is thus given by
 \begin{equation}
\left(1 +\frac{1}{2} F_{1/2}(\tau_t)\right)
\frac{\alpha_s}{8\pi}\frac{r}{\Lambda_r}G_{\mu\nu}G^{\mu\nu},
\label{inducedgluon}
\end{equation}
while for the photon
\begin{equation}
\left( \frac{7}{3}+ F_1(\tau_W) + \frac{4}{3} F_{1/2}(\tau_t) \right)
\frac{\alpha_{EM}}{8\pi}\frac{r}{\Lambda_r}F_{\mu\nu}F^{\mu\nu}.
\label{inducedphoton}
\end{equation}
 The final coupling is then obtained as a
 sum of the bulk contribution (\ref{masslessGBcoupl}) and the IR brane
 localized terms (\ref{inducedgluon}) for the gluon or
 (\ref{inducedphoton}) for the photon, given by
\begin{equation}
-\frac{r}{4\Lambda_r} \left( \frac{R}{g_5^2}+\frac{(b_{IR}-\sum_i
 \kappa_i F_i(\tau_i))}{8\pi^2} \right) g^2 F_{\mu\nu}F^{\mu\nu},
\label{fullcoup}
\end{equation}
where $\kappa^3_t=1/2$, $\kappa^{EM}_{t}=4/3$, $\kappa^{EM}_{W}=1$, and $g$ is the one loop corrected 4D gauge coupling, which can be
approximately expressed as:
\begin{equation}
\frac{1}{g^2 (q)} = \frac{R \log R'/R}{g_5^2} + \tau^{(0)}_{\rm UV} +
\tau^{(0)}_{\rm IR} - \frac{b_{\rm UV}}{8 \pi^2} \log R'/R,
\label{4Dgcoup}
\end{equation}
where we have neglected terms with small logs of order $\frac{1}{8
\pi^2} \log \frac{1}{q R'}$. Here $b_{UV}$ contains the beta function
coefficients due to the
rest of the SM fields, either localized on the UV brane or in
the bulk: $b_{UV}^{EM}= -4/3$ and $b_{UV}^3= 8$.

We can solve Eq.~(\ref{4Dgcoup}) for $R/g_5^2$, and plug into
Eq.~(\ref{fullcoup}) to obtain the final result:
\begin{equation}
-\frac{r}{4 \Lambda_r \log R'/R} \left[ 1 - 4 \pi \alpha \left(
  \tau^{(0)}_{\rm UV} + \tau^{(0)}_{\rm IR} \right) + \frac{\alpha}{2 \pi} \left(
 b -\sum_i
 \kappa_i F_i(\tau_i) \right)  \log R'/R  \right] F_{\mu \nu} F^{\mu
  \nu},
\label{fincoup}
\end{equation}
where $b$ is the total beta-function coefficient, including Planck, TeV
brane localized, and flat fields: $b = b_{\rm UV} + b_{\rm
  IR}$.  Note that in this final formula it does not matter how a
Plank or TeV brane localized (or flat) field contributes to the
running; only the total beta function is relevant (and it is fixed
by the SM value). Also, the loop effects of the massive $W$ will be
decoupling since the total $b_{UV}^{EM}+b_{IR}^{EM}=7$ cancels against
the triangle diagram contribution of the massive $W$.  The same is true for the
contributions of the massive top quark.

\section{The CFT Interpretation of the radion couplings}
\label{sec:CFT}
\setcounter{equation}{0}
\setcounter{footnote}{0}

The AdS/CFT correspondence can be extended to the two-brane RS type
models under consideration here~\cite{CFT}. The main new ingredient
(vs. the 4D interpretation of the infinite AdS space) is that the
two branes correspond to breaking of conformal invariance: the UV
brane acts as a UV cutoff, while the IR brane corresponds to
spontaneous breaking of conformal invariance. The radion appears as
a normalizable mode only when the IR brane is included, and without
stabilization it would be a massless field. Thus it is natural to
identify it with the Goldstone boson corresponding to the
spontaneous breaking of conformal invariance due to the appearance
of the TeV brane~\cite{CFT}. This identification also suggests that
one can also give a simple CFT interpretation of the radion
couplings explicitly calculated in the previous sections. We will
assume that there are two types of fields: those localized on the
Planck brane and those on the TeV brane. From the CFT point of view,
the Planck brane fields correspond to elementary fields which are
only indirectly coupled to the CFT (either through gauge or
gravitational interactions), while the TeV brane localized fields
are composites of the CFT. The only special field not in this class
is a gauge boson zero mode. This corresponds to a weakly gauged
global symmetry of the CFT, and the flat zero mode can actually be
thought of as a mixture of an elementary and a composite spin one
field (analogous to the $\gamma - \rho$ mixing which is present in
QCD)~\cite{CFT,CFTgauge}.

Non-derivative Goldstone couplings are usually a consequence of
explicit symmetry breaking.  Generally, such couplings are
given by
\begin{equation}
\frac{r}{f} \partial_\mu J^\mu,
\end{equation}
where $J^\mu$ is the global current whose spontaneous breaking is
leading to the appearance of the Goldstone, and $f$ is the symmetry
breaking scale. In the case of dilatation symmetries $J_\mu =
\theta_{\mu\nu}x^\nu$, where $\theta_{\mu\nu}$ is the symmetric
energy-momentum tensor.  The general form of the radion coupling is
thus expected to be of the form
\begin{equation}
\frac{r}{f} \theta_\mu^\mu \ .
\end{equation}
In order to find the coupling to massive CFT composite modes (fields
peaked towards the TeV brane), we just take the general expression
for $\theta_\mu^\mu$, which for massive spin 1/2 fields is $m
\bar{\psi} \psi$ while for massive spin 1 fields it is $M_A^2 A_\mu
A^\mu$. If we identify $f$ with $\Lambda_r$, we obtain the leading couplings to massive fields
as in Eqs. (\ref{topquarkcoupl}) and (\ref{WZcoupl}).

An interesting special case is the
coupling to light fermions, which are localized on the Planck brane rather
than the TeV brane. The explicit calculation shows that the coupling
is given by (\ref{lightfermcoupl}). The appearance of the factor $c_R-c_L$
might seem mysterious, however it is perfectly clear from the CFT picture.
The CFT interpretation of a Planck-brane localized left handed fermion
zero mode
$\chi_L$ is that an elementary fermion is mixed with a dimension $2+c_L$
composite operator ${\cal O}_L$ in the CFT~\cite{CFTfermions}. Similarly the
RH fermion zero mode is described as an elementary fermion $\psi_R$ mixed
with a $2-c_R$ dimensional CFT operator ${\cal O}_R$. The mass term for the
fermion in the 5D picture arises from the TeV brane, and in the CFT it
is simply
an operator~\footnote{We show in Appendix~\ref{app:fermionCFT} that
  this indeed leads to the
fermion mass formula (\ref{eq:fermmass}).}
\begin{equation}
{\cal L}_{mass} =\lambda {\cal O}_L {\cal O}_R.
\end{equation}
The dimension of this operator is $4+c_L-c_R$, thus the coefficient has scaling
dimension $c_R-c_L$. A simple way of calculating the coupling of the Goldstone
is by writing down the non-linear $\sigma$-model term for this operator. This can be achieved
by compensating the scaling dimension of the above operator  with a factor of
$e^{(c_R-c_L)r /f}$.  Therefore the radion coupling should be given by expanding
\begin{equation}
\lambda {\cal O}_L {\cal O}_R e^{(c_R-c_L) r/f},
\end{equation}
which to linear order is just
\begin{equation}
\lambda \left(1+\frac{c_R-c_L}{f} r \right) {\cal O}_L {\cal O}_R.
\end{equation}
After rotating to the mass eigenstates we find the relevant term to be
\begin{equation}
-m\left(1+\frac{c_R-c_L}{f} r\right) \bar{\psi}_L \bar{\psi}_R,
\end{equation}
in agreement with (\ref{lightfermcoupl}).

In order to obtain the coupling to
the massless gauge fields, we need to understand the anomalous
contribution to $\theta_\mu^\mu$ from massless gauge fields, since in the
CFT language the entire coupling will come from this source. We
know that the trace anomaly is generically given by
\begin{equation}
\theta_\mu^\mu = -\frac{b \alpha}{8\pi} F_{\mu\nu}F^{\mu\nu},
\end{equation}
where $b$ is the $\beta$-function coefficient leading to anomalous
scale-invariance violations and $\alpha = \frac{g^2}{4\pi}$.

\begin{figure}
\centerline{\includegraphics[width=.3\hsize]{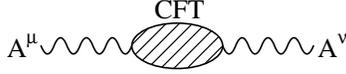}}
\caption{The free gauge propagator of the CFT} \label{CFTdiagram}
\end{figure}

In the case of the radion, we need to take into account that loops of
CFT fields will give rise to a contribution to the 2-point function
of the fundamental gauge fields.  This is the usual effect in the
running of the coupling due to the bulk given in
Fig.~\ref{CFTdiagram}, corresponding to an effective $\beta$-function from
the CFT: 
\begin{equation}
b_{CFT} =-\frac{8\pi^2 R}{g_5^2}.
\end{equation}
It is only the effect of CFT loops that directly contribute to
the coupling of the radion, since elementary fields localized on the
Planck brane do not feel the effect of spontaneous breaking of scale
invariance.  However, they will have an indirect effect due to the
matching of the 4D value of the gauge coupling $\alpha$. The
relevant formula for the 4D gauge coupling at a low-scale is given
by
\begin{equation}
\frac{1}{g^2}=\frac{R \log \frac{R'}{R}}{g_5^2}+\tau_{UV}+\tau_{IR},
\end{equation}
where $\tau_{UV,IR}$ are the brane localized kinetic terms on the
two branes. Thus we again obtain the coupling
\begin{equation}
-\frac{R}{R \log{\frac{R'}{R}\ +(\tau_{UV}+\tau_{IR})g_5^2}}
\frac{r}{4\Lambda_r}
F_{\mu\nu} F^{\mu\nu},
\end{equation}
which is identical to the bulk 5D result in (\ref{masslessGBcoupl}). However,
we have seen that in the 5D picture there is also a contribution
to the coupling from the IR brane localized trace anomaly and triangle
diagrams. In the CFT this effect due to light composites is somewhat subtle. One simple
way of thinking about this is to say that part of the CFT states go into forming massive
composites, but another part of the CFT will go into forming massless composites. Since the
massless composites do not feel the effects of spontaneously broken conformal invariance,
these states will not contribute to the coupling of the radion. Thus their contribution actually
has to be {\it subtracted} from the entire CFT contribution. Assuming that the part of the CFT that
forms the light composites has a $\beta$-function coefficient $b^{IR}$, the corrected coupling will be given
by
\begin{equation}
\frac{r}{4 \Lambda_r} \frac{(b^{CFT}-b^{IR})\alpha}{2\pi} F_{\mu\nu}F^{\mu\nu}=
\frac{r}{4 \Lambda_r} \left[ -  4 \pi \alpha \frac{R}{g_5^2}
-\frac{\alpha}{2 \pi} b^{IR}\right] F_{\mu\nu}F^{\mu\nu} .
\label{CFTfull}
\end{equation}
Here we again identify $b^{IR}$ with the effects of a composite Dirac top quark,
the left handed $b$ quark, and a composite Higgs.  Then, as before, 
$b^{IR}_3=-1$ for the gluon and $b^{IR}_1=-7/3$ for the photon.
To this we add the effects of the triangle diagrams from the light composites, which include
the top quark and the massive gauge $W$ (containing the eaten components of the composite Higgs). This way
for the effects of the light composites we get the expressions in
Eq.~(\ref{inducedgluon}) for the gluon and Eq.~(\ref{inducedphoton}) for the photon,
in agreement with the 5D results.

Perhaps the simplest and most unambigous way to identify the radion coupling to the light gauge
bosons is by considering the $R'$ dependence of the matching relation for the couplings.\footnote{We thank
Kaustubh Agashe for suggesting this method to us.} The expression of the 4D effective action for the
massless gauge fields is given by (before going to canonical normalization of the gauge field):
\begin{equation}
-\frac{1}{4g^2} F_{\mu\nu}F^{\mu\nu},
\end{equation}
where the full matching of the coupling is
\begin{equation}
\frac{1}{g^2(q)}= \frac{R\log \frac{R'}{R}}{g_5^2} +\tau_{UV}^{(0)} -\frac{b^{UV}}{8\pi^2}
\log \left(\frac{1}{Rq}\right)
+\tau_{IR}^{(0)} -\frac{b^{IR}}{8\pi^2}\log \left( \frac{1}{R'q}\right).
\label{fullmatching}
\end{equation}
Here $b^{UV}$ includes the light fields localized on the UV brane plus the effects of the massless
gauge boson zero modes, while $b^{IR}$ only contains the IR localized fields (Dirac top, left handed bottom, and Higgs).
Since the radion in the 4D effective theory can be interpreted as the fluctuation of the IR scale $R'$, the coupling to
the radion can be obtained by the replacement $R'\to R'(1+r/\Lambda_r )$ in the above formula, from which we
immediately obtain (after going back to canonically normalized 4D gauge field) the coupling to be
\begin{equation}
-\frac{r}{4 \Lambda_r} \left(\frac{R}{g_5^2}+\frac{b^{IR}}{8\pi^2}\right)g^2 F_{\mu\nu}^2,
\end{equation}
in agreement with (\ref{CFTfull}).

We also find that the decoupling of states such as the
$W$ and top quark occur in a rather simple way with this 
method, without calculating any triangle diagrams.  When $q$ is less
than the mass, $m_i$, of a species which contributes to the running,
Eq.~(\ref{fullmatching}) is modified such that the logarithmic running
from this field ends at the scale $m_i$ rather than $q$.  These
effects are incorporated by adding the following terms to Eq.~(\ref{fullmatching})
\begin{equation}
\sum_i \frac{b_H^i}{8 \pi^2} \log \frac{m_i}{q} \approx 
\frac{b_H(q)}{8\pi^2} \log \frac{1}{q R'},
\label{matchingwmass}
\end{equation}
where we define $b_H^i$ as the beta function
coefficients due to species with mass greater than the momentum transfer
$q$, and $b_H(q)$ as their sum. 
Using the same replacement as before, $R'\rightarrow~R'(1+r/\Lambda_r)$,
the total radion coupling is then found to be
\begin{equation}
-\frac{r}{4 \Lambda_r}
 \left(\frac{R}{g_5^2}+\frac{b^{IR}-b_H(q)}{8\pi^2}\right)g^2
 F_{\mu\nu}^2.
\label{decouple}
\end{equation}
Solving Eq.~(\ref{fullmatching}) for $R/g_5^2$ (neglecting subleading logs,
as usual), and plugging into
Eq.~(\ref{decouple}) then gives the final result
\begin{equation}
-\frac{r}{4 \Lambda_r \log R'/R} \left[ 1 - 4 \pi \alpha
  (\tau_{UV}^{(0)}+\tau_{IR}^{(0)})  + \frac{\alpha}{2 \pi} (b - b_H(q)) \log R/R'
   \right] F_{\mu\nu} F^{\mu\nu},
\end{equation}
where $b$ is the total beta function coefficient from all SM fields,
and decoupling is manifest.  Up to threshold effects, this is in
complete agreement with Eq.~(\ref{fincoup}).

Further corrections are expected to be suppressed by $1/N$ in the CFT.
Since the usual identification is $1/N = g_5^2/(16\pi^2 R) =
g_4^2\left( \log R'/R \right)/(16 \pi^2)$, these $1/N$ corrections are
at  most ten percent (if $\log R'/R \sim 30$) for the photon, and at
most 25 percent for the gluon.  While for a precision measurement
these effects would need to be taken into account, here we will be
satisfied with obtaining the leading order estimates.

\section{Radion branching fractions}
\setcounter{equation}{0}
\setcounter{footnote}{0}

In this section we present our results for the branching fractions of
the radion, assuming as we did throughout this paper that the Higgs
is completely localized on the IR brane.  We also assume that there is no
Higgs-curvature coupling, such that Higgs-radion mixing is absent.  It
is worth noting that this coupling is actually expected to be small
in the case that the Higgs is an approximate Goldstone boson of
a spontaneously broken global symmetry~\cite{GRW} (as is the case in holographic
composite Higgs models~\cite{holoHiggs}).  Such mixing is also
generically small in Higgsless models of EWSB~\cite{higgsless}.
Therefore this assumption is a reasonable one in many realistic warped
space models of EWSB.

We first note that radion production at the LHC could be substantial due to the
fact that the
branching fraction of the radion to two gluons could 
be enhanced by as much as a factor of $10$ (for $\Lambda_r = 1$~TeV)
in comparison with the Higgs branching fraction to gluons.
The enhancement is due to the fact that the radion couples to massless
gauge bosons through the conformal anomaly, which is rather large for
QCD.

A promising signal for the radion is in the $\gamma\gamma$ decay
channel.  The tree level coupling of the radion to massless gauge bosons, which
is absent in the RS1 model, would naively enhance the branching ratio to photons by a large
factor.  However, as can be seen in Eq.~(\ref{fincoup}), the tree level coupling
cancels with the loop-level terms since the QED beta
function coefficient is negative.   

If we fix $R$ and $R'$, the only remaining free parameter entering the coupling
to a massless gauge boson is given by the sum of the bare localized
kinetic terms for that field:  $\tau^{(0)}_{UV}+\tau^{(0)}_{IR}$.
This sum can be bounded by requiring perturbativity of the low-energy
theory. We can, for example, require that the 5D NDA cutoff scale is
at least 10 TeV:
\begin{equation}
\frac{24 \pi^3}{g_5^2} \frac{R}{R'}=\Lambda_{\rm IR}\geq 10 {\rm TeV},
\end{equation}
from which we get (neglecting the small logarithm from IR brane
running) an upper bound of 
\begin{equation}
\tau^{(0)}_{\rm UV}+\tau^{(0)}_{\rm IR} \leq \frac{1}{g^2} -
\Lambda_{\rm IR} R'\frac{\log R'/R}{24 \pi^3}  + 
\frac{b_{\rm UV}}{8\pi^2} \log R'/R.
\label{pertbnd}
\end{equation}

\begin{figure}[t]
\includegraphics[width=1.0\hsize]{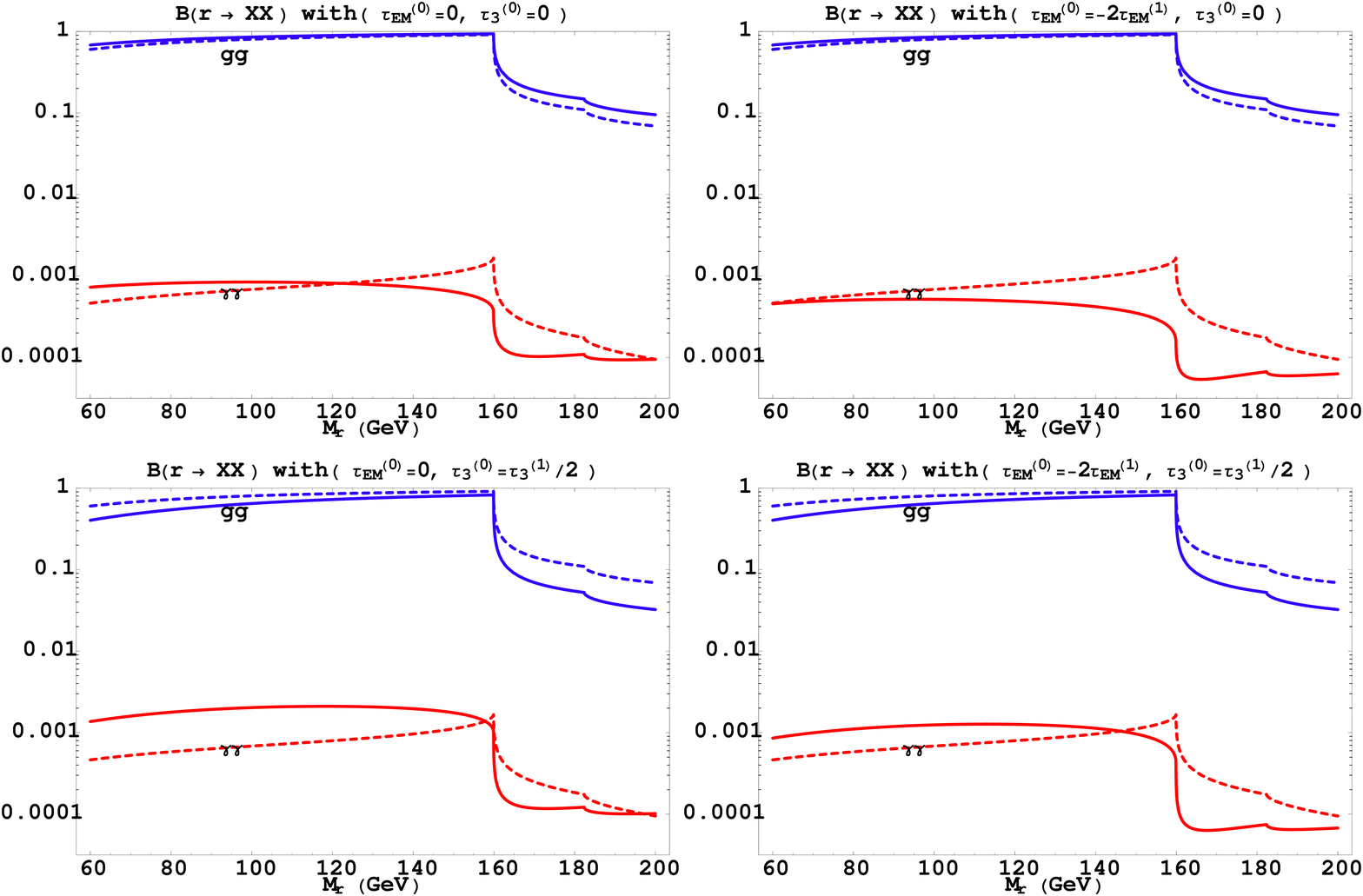}
\caption{In these plots, we show the branching fractions
of
  the radion into gluons and photons (the solid lines), comparing with
  RS1 scenario (the dashed curves). For each graph, the solid curves
represent the branching fractions in the
presence of different combinations of tree level brane localized
kinetic terms for the gluon and photon.  The magnitudes of
the localized terms are given on the top of every plot
individually in units of the one-loop corrections to the $\tau^{\rm
  UV}_i$. We have set $\Lambda_r=2$~TeV, corresponding to $1/R'=816$~GeV.
 } \label{branchingfractions1}
\end{figure}

For our analysis in the remaining part of this paper, we set
$\tau_{IR}^{(0)}=0$, assuming the tree level IR brane localized
kinetic terms to be negligibly small. In
Figure~\ref{branchingfractions1} we show how the branching fractions 
of the radion into gluons and photons are modified in the presence of
non-vanishing tree level brane localized kinetic terms.  In all of the
plots, we assume that the Higgs mass is $115$~GeV.  In our
figures, we express the
bare BKTs in units of the one-loop corrections to the UV brane
localized kinetic terms:
\begin{equation}
\tau^{(1)}_i=\frac{b^{\rm UV}_i}{8\pi^2} \log{\frac{R'}{R}}.
\end{equation}
In order to avoid the possibility of ghosts in the spectrum, we only
consider positive tree level BKTs.  We assume that
$t_L$, $t_R$, $b_L$, and the Higgs are on the IR brane, and thus
$b^{\rm UV}_{\rm EM} = -4/3$, and $b^{\rm UV}_3 = 8$.  The perturbativity constraint in
Eq.~(\ref{pertbnd}) then requires that the gluon BKT not
exceed about $\tau^{(0)}_3 < (0.9 - 1.1) \cdot \tau_3^{(1)}$, while for the photon,
$\tau^{(0)}_{\rm EM} < -( 14 - 16 ) \cdot \tau_{\rm EM}^{(1)}$, where
we have varied $\Lambda_{\rm IR}$ between $2$ and $20$~TeV.

Note that in the absence of sizable tree level brane localized
kinetic terms, the branching fractions for the massless gauge bosons are
slightly larger in comparison with the RS1 case. 
The effect of introducing significant tree level brane localized
kinetic terms is to reduce the branching fractions for the
corresponding gauge boson.  This is due to the fact that the effect of
positive BKTs cancels against the term
that scales as $1/\log R'/R$.
In the presence of a tree level brane localized kinetic term for the
gluon only, the branching fraction to photons is enhanced. The
reason for this is that the decay to gluons is the dominant mode.
Once a BKT for the gluon is added, the partial width to gluons
decreases, thereby increasing the branching fractions to subdominant channels.

\begin{figure}[t]
\includegraphics[width=1.0\hsize]{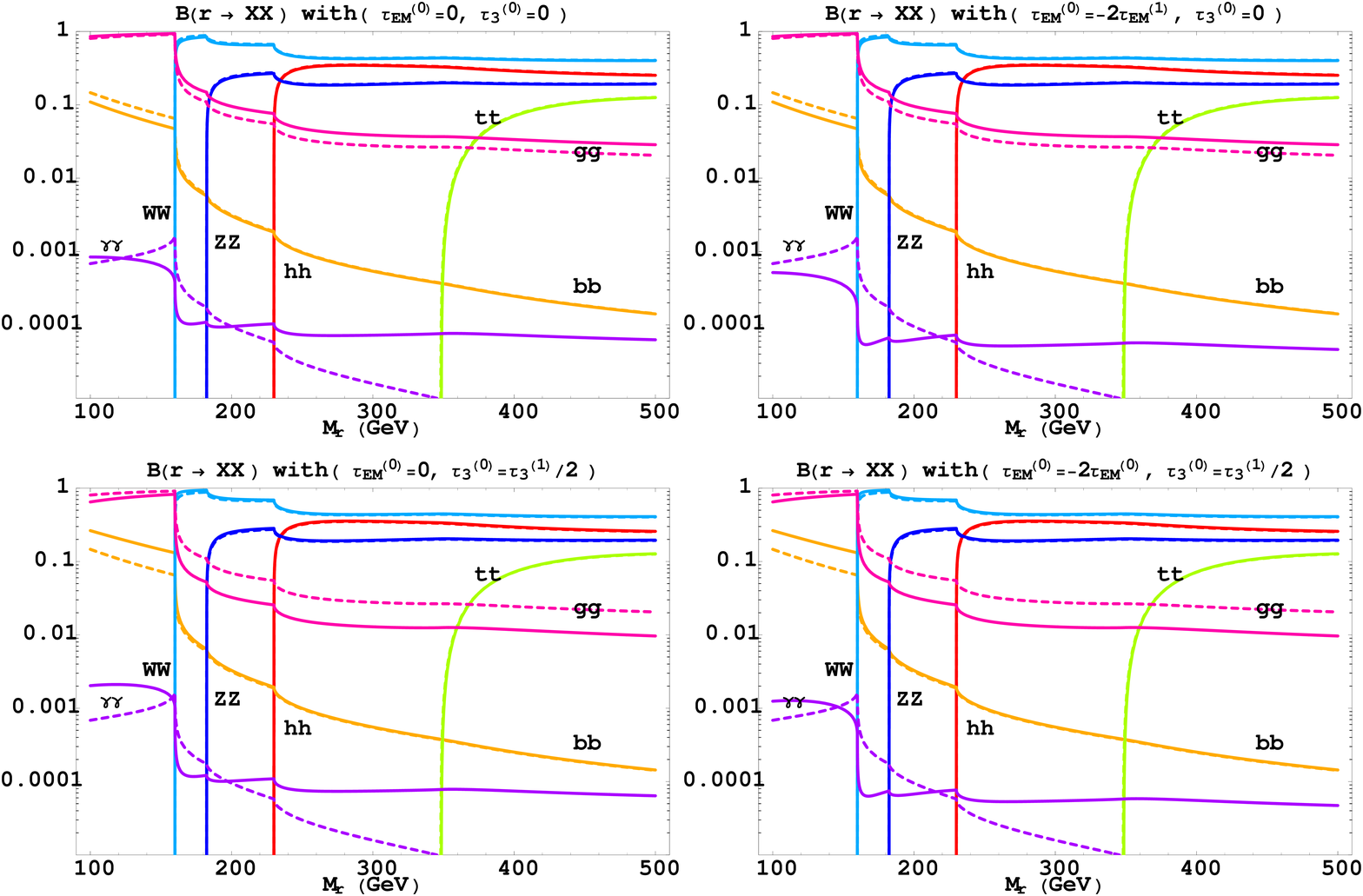}
\caption{In this plot, we show the branching fractions of the radion
  into on-shell final states.
  Again, the dashed curves represent the branching fractions in the
  RS1 scenario. We show the effect of introducing tree level brane localized
  kinetic terms for the photon and gluon, choosing different
  combinations for each graph. We set $\Lambda_r=2$~TeV, corresponding
  to an IR brane scale of $1/R'=816$~GeV. 
} \label{highMr}
\end{figure}

In Figure~\ref{highMr}, we demonstrate how the 
branching fractions of the radion into all SM particles change in
the presence of tree level brane localized gauge boson kinetic
terms.  The branching fractions to massive gauge bosons,
Higgs bosons, and top quarks are not significantly different compared
to the original RS1 scenario.
Note however that the branching fraction to gluons (and thus the
production rate) for larger radion masses still differs from the values
found in the brane localized SM.

\section{Discovery potential at the LHC}
\setcounter{equation}{0}
\setcounter{footnote}{0}

In this section we discuss the radion discovery potential at the
LHC. We show plots of the ratio of the discovery significance of the 
radion in the $gg\rightarrow r\rightarrow ZZ\rightarrow 4l$ and
$gg\rightarrow r\rightarrow \gamma\gamma$ channels and those
of a SM Higgs boson with the same mass. An approximate formula for the
ratio of the significance 
of potential radion discovery compared with a SM Higgs boson discovery
of the same mass for the
$gg\rightarrow r\rightarrow \gamma\gamma$ channel is given in~\cite{GRW}:
\begin{equation}
R_S^{\gamma
\gamma}\equiv\frac{S(r)}{S(h_{SM})}=\frac{\Gamma(r\rightarrow g
g)\,B(r\rightarrow \gamma\gamma))}{\Gamma(h_{SM}\rightarrow g
g)\,B(h_{SM}\rightarrow
\gamma\gamma))}\sqrt{\frac{max(\Gamma_{tot}(h_{SM}),\Delta M_{\gamma
\gamma})}{max(\Gamma_{tot}(r),\Delta M_{\gamma \gamma})}},
\end{equation}
with a similar formula applying for the significance ratio in the $r
\rightarrow ZZ$ discovery channel.

The factor inside the square root measures the ratio of the relative effective
total widths of the Higgs and radion as they would appear in the
detector.  For smaller widths, the signal to background ratio is
higher, although this effect is limited by the detector resolution for
diphoton (or 4 lepton) invariant masses.  
If the total width is smaller than the energy resolution, the
entire signal is contained in a single bin, and one then needs
to consider the background over that entire region (rather than only over the
energy range given by the width of the decaying particle).

\begin{figure}[t]
\includegraphics[width=1.0\hsize]{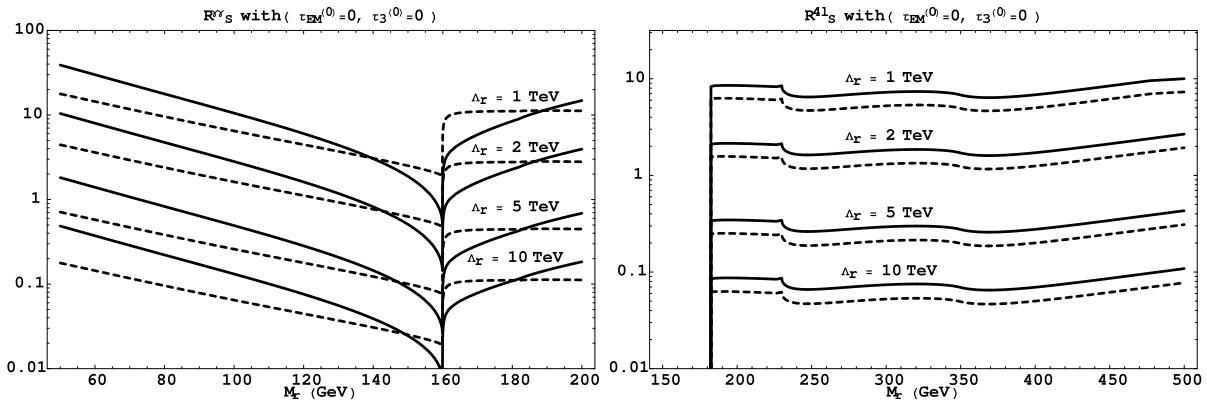}
\caption{In the plot on the left,we show a comparison of the ratio of
  discovery significance for a radion vs. a Higgs of the same mass,
  $R^{\gamma\gamma}_S$, with the scenario where the SM fields are all
  localized on the IR 
brane (the dashed curves).  In the plot on the right we show the ratio
  of discovery significance $R^{4l}_S$. We assume that there are no
  tree level brane localized kinetic terms for the gluon or
  photon. For the displayed values of $\Lambda_r$, the corresponding
  values of $1/R'$ are $408$, $816$, $2041$, and $4082$~GeV. 
} \label{significance}
\end{figure}

In Fig.~\ref{significance} we plot $R^{\gamma\gamma}_{S}$ and
$R^{4l}_{S}$ in the case that there are no tree level brane localized kinetic
terms for either the gluon or photon. We find that for low values of
$\Lambda_r$, the ratio $R^{\gamma\gamma}_{S}$ is always greater than
one, implying that one is more likely to find a radion of this mass
than a Higgs of the same mass.  For some values of the radion mass,
$R^{\gamma\gamma}_{S}$, is enhanced compared to the case with all fields
localized on the IR brane, up to a factor of 3 for large values of
$\Lambda_r$.  In the $r \rightarrow 4l$ channel, there is a generic
enhancement of the discovery potential in comparison with the IR brane
localized SM scenario due to the larger $r \rightarrow gg$ branching fraction.

\begin{figure}[t]
\includegraphics[width=1.0\hsize]{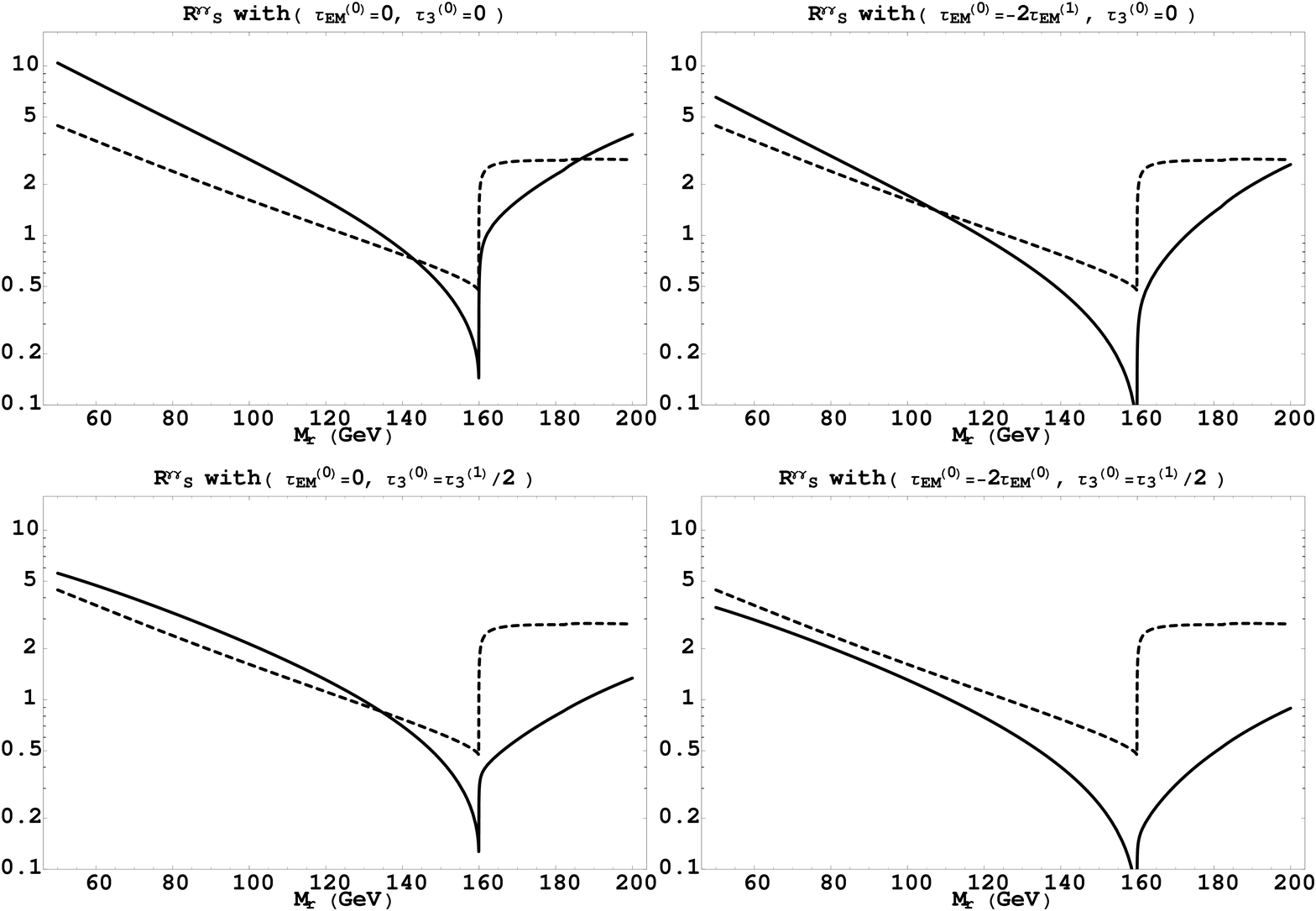}
\caption{In this figure, we show the ratio of the discovery
significance $R^{\gamma\gamma}_{S}$ with $\Lambda_r=2$~TeV for various
combinations of tree level brane localized kinetic terms for the
photon and gluon.  We have set $\Lambda_r=2$~TeV, corresponding to an
IR brane with $1/R'=816$~GeV.   Again, the dashed curves represent the
signal significance in the original RS1 model. 
} \label{discoverysignificance}
\end{figure}

In Fig.~\ref{discoverysignificance} we  plot $R^{\gamma\gamma}_{S}$
for different combinations of tree level brane localized kinetic terms
for both the gluon and the photon, taking $\Lambda_r=2$~TeV.
One can see that turning on positive BKTs
generically reduces the potential radion signal in the diphoton
channel, and that the signal can even be reduced compared to the 
traditional RS1 model.

\begin{figure}[t]
\includegraphics[width=1.0\hsize]{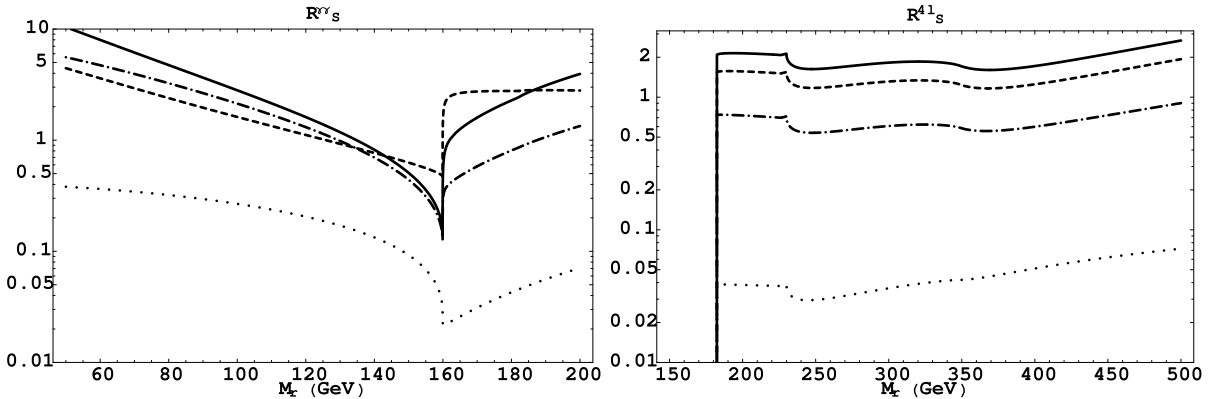}
\caption{In this figure, we compare the discovery significance in
  $\gamma\gamma$ and
  $4l$ in the presence of a UV brane localized kinetic term for
  the gluon only: $\tau_{EM}^{(0)}=0$.  The curves that are
  (dashed, solid, dashed-dotted, dotted) correspond to
  the RS1 scenario, $\tau_3^{(0)} = 0$,
  $\tau_3^{(0)} = \tau_3^{(1)}/2$, and $\tau_3^{(0)} = \tau_3^{(1)}$, respectively.
  This last value saturates the  
  perturbativity bound in Eq.~(\ref{pertbnd}) for most values of $\Lambda_{IR}$.
} \label{significance4L}
\end{figure}

Finally, in Fig.~\ref{significance4L} we plot $R^{\gamma\gamma}_{S}$ and $R^{4l}_{S}$ for
different values of the tree level brane localized kinetic term for
the gluon, taking $\Lambda_r = 2$~TeV, and $\tau_{EM}^{(0)} =0$. 
The signal significance $R^{4l}_{S}$ is insensitive to brane localized 
kinetic terms for the photon.  We choose three values for the gluon UV
BKT, $\tau_3^{(0)}=0$, $\tau_3^{(1)}/2$, and $\tau_3^{(1)}$.  The
first two are well within the perturbativity constraint in
Eq.~(\ref{pertbnd}), while the last choice saturates it.  For the 
last case, the branching fraction to gluons is quite suppressed, as
the large QCD conformal anomaly contribution is nearly canceled by the
gluon BKT.  The production cross section is therefore strongly
quenched.  We find in this last case that the signal significance can
be significantly smaller (by a factor of 10) compared to the
traditional RS1 model. 


For all cases we have considered, there is a small region where
$R^{\gamma\gamma}_{S}$ is sharply suppressed, corresponding to the
$\sim 160$~GeV
threshold at which the radion decay to $WW$ turns on, and the
normalized signal significance of $gg \rightarrow r \rightarrow
\gamma\gamma$ drops sharply.  In addition,
there is a sharp step when $R^{4l}_{S}$ becomes well defined (when the
on-shell decays $r,h \rightarrow ZZ$ are kinematically allowed).  We note that
this behavior is softened if one takes into account decay
channels which pass through off-shell $W$ or $Z$ bosons, and that the
gap between the $WW$ and $ZZ$ thresholds is in fact covered by the $4l$
signal with an off-shell intermediate $Z$-boson.

In summary, we have found that there are large regions of parameter space where
one is as likely to find a radion as a SM Higgs of the same mass.  With
$100$~fb$^{-1}$ of data, it is projected that a SM Higgs would be
discovered with a confidence level of between $10$ and $13$ sigma for
Higgs masses between $100$ and $1000$~GeV.  Therefore if the solution of the SM
hierarchy problem involves warped extra dimensions, it is not
unlikely that we will discover a radion at the LHC.  When we find a
new particle with Higgs-like signatures, it becomes a matter of deciding
whether or not we have found ``The Higgs.''  In the large mass region
(when the resonance is above about $230$~GeV), a SM Higgs boson would have a
width which is resolvable by the detector, while the radion width is
much narrower.  A width measurement could exclude a conventional SM
Higgs boson as a candidate for this new state.  The smaller mass
range deserves further study, as a width measurement will not suffice,
making discrimination more difficult.  Precision measurements (such as
those that would be performed at a linear collider), coupled
with a discovery of KK towers for the SM gauge and matter fields would
provide further evidence helping to confirm the discovery of a RS radion. 

\section{Conclusions}

We have investigated the phenomenology of the RS radion in models
where the SM fields propagate in the bulk as opposed to being confined
to the IR brane in the traditional RS1 model. The motivation for this is
that most realistic models of EWSB utilize such a setup. 
We have calculated the radion couplings to the SM fields in detail, and we have
given a CFT interpretation for all of these interactions. We have paid
special attention to the couplings to massless gauge bosons, since
these are phenomenologically the most important ones, and since the
one-loop effects are significant. We then compared the radion discovery
potential of 
the LHC to that for a SM Higgs, and also to the radion of the
traditional RS1 model. We have found that the $\gamma\gamma$ signal
is enhanced over the SM Higgs case for some reasonable values of
the RS scale.  However, this signal depends quite sensitively on the
values of the brane localized kinetic terms for the massless gauge
fields. If those are sizeable, the significance of the signal can
decrease, becoming smaller than in the traditional RS1 case.
Finally, we have shown that for the region of larger radion masses where
the $\gamma\gamma$ signal is no longer significant, one can use the 4
lepton signal to look for the radion.  If a new Higgs-like state is
discovered at the LHC, a width measurement could rule out a
conventional Higgs boson, however further discoveries involving KK
modes and precision studies of the scalar sector would be
necessary to provide convincing evidence for a RS radion. 

\section*{Acknowledgments}

We are extremely grateful to Kaustubh Agashe for many useful discussions
of the CFT interpretation of the radion and for clarifying
discussions about the running and matching of the 5D gauge coupling,
and its application to calculating the radion coupling to massless
fields. We also thank Brian Batell, Matthias Neubert, Matt Reece, Jose
Santiago, and Manuel Toharia for useful discussions. The research of
C.C. is supported in 
part by the DOE OJI grant DE-FG02-01ER41206 and in part by the NSF
grant PHY-0355005. The research of J.H. is supported by Fermilab,
which is operated by Fermi Research Alliance, LLC under Contract
No. DE-AC02-07CH11359 with the US DOE.  The
research of S.J.L. is supported in part by 
the NSF grant PHY-0355005.

\section*{Appendix}
\setcounter{section}{0}
\renewcommand{\thesection}{\Alph{section}}

\section{The cancellation of the boundary couplings with fermions}
\label{app:boundcancel}
\renewcommand{\theequation}{A.\arabic{equation}}
\setcounter{equation}{0}

The actual construction for fermion masses for the bulk fermions involves
adding boundary mass terms for the fermions of the form
\begin{equation}
-\int d^4x\sqrt{g^{ind}} M_D R' (\psi_R \chi_L +h.c.)
\label{boundmass}
\end{equation}
Note, that since $\psi ,\chi$ are bulk fermions, the actual mass parameter is a
dimensionless ${\cal O}(1)$ coupling which we denote for convenience denote by $M_D R'$.
Expanding this in terms of the radion field we obtain the boundary coupling
\begin{equation}
\int d^4x \left( \frac{R}{R'}\right)^4 4F M_D R' (\psi_R\chi_L+h.c.).
\label{boundarycoupling}
\end{equation}
However, there is an additional localized coupling term from the discontinuity
of the fermion wave functions around the TeV brane. The simplest way to see
the emergence of this term is by assuming that the actual BC's at the TeV brane
are $\chi_R=\psi_L=0$ irrespectively of the boundary mass term, and then add the boundary
mass a distance $\epsilon$ away from the TeV brane. The effect of this boundary mass
(\ref{boundmass}) will then be to induce a discontinuity in the fermion wave functions
\begin{equation}
[\chi_R]\equiv -\chi_R|_{R'-\epsilon}+\chi_R|_{R'}, \ \ [\psi_L]\equiv -\psi_L|_{R'-\epsilon}+\psi_L|_{R'}.
\end{equation}
Here $\chi_R|_{R'}$ and $\psi_L|_{R'}$ are still fixed to zero, and the limiting values
at distance $\epsilon$ are the fields appearing in the boundary conditions for the fermions.
Then the bulk action will also give a contribution to the radion coupling with
magnitude
\begin{equation}
\int d^4x \left( \frac{R}{R'}\right)^4 2 F (\psi_R [\chi_R]-[\psi_L]\chi_L +h.c.).
\label{bulkjump}
\end{equation}
However, evaluating the discontinuities using the fermion BC's
we find that (\ref{bulkjump}) exactly cancels
(\ref{boundarycoupling}).

\section{Fermion masses from CFT}
\label{app:fermionCFT}
\renewcommand{\theequation}{B.\arabic{equation}}
\setcounter{equation}{0}

We have seen from the explicit calculation that the approximate expression for the fermion
masses is given by (\ref{eq:fermmass}). Here we present a simple derivation of this result
from the CFT point of view. Most of the elements of this are already implicitly
contained in \cite{CFTfermions}. As explained in Section~\ref{sec:CFT}, the fermion
zero mode (before EWSB)  is interpreted as a mixture of an
elementary fermion $\chi_L$ with an with a composite operator ${\cal O}_L$ (and similarly
the right handed $\psi_R$ is mixed with ${\cal O}_R$). The mixing among these fields is
determined by the wave function of the actual zero mode on the IR brane:
\begin{equation}
\omega_{R,L}= \sqrt{\frac{1\pm 2c_{R,L}}{1-(\frac{R}{R'})^{1\pm
2c_{R,L}}}}
\end{equation}
Thus the mixture that is the zero mode can be identified with (assuming $\omega_{L,R}\ll 1$)
\begin{equation}
\chi_{light}=\chi_L -\omega_L {\cal O}_L\ , \ \psi_{light}=\psi_R -\omega_R {\cal O}_R\ .
\end{equation}
As discussed in Sec.~\ref{sec:CFT} the electroweak symmetry breaking mass term is
given by $M_D {\cal O}_L {\cal O}_R$. Expressing ${\cal O}_{L,R}$ in terms of the
light fields we find that the mass for the light fields is given by
\begin{equation}
M^2= M_D^2 \omega_L^2 \omega_R^2 = M_D^2 \frac{(1-2c_L)(1+2c_R)}{(1-(\frac{R}{R'})^{1-2 c_L})
(1-(\frac{R}{R'})^{1+2 c_R})}.
\end{equation}
This agrees with the expression (\ref{eq:fermmass}). For $c_L>1/2, c_R<-1/2$ this simplifies to
\begin{equation}
\frac{\lambda}{R} \sqrt{(2 c_L-1)(-1-2c_R)}\left( \frac{R'}{R}\right)^{c_R-c_L},
\end{equation}
where $\lambda = M_D R'$ is the fundamental parameter that we are adding to the theory. One can read off
the radion coupling from this form of the mass formula by substituting the $R'$ dependence with the radion.

\section{Cancellation of linear divergences}
\label{app:Lindiv}
\renewcommand{\theequation}{C.\arabic{equation}}
\setcounter{equation}{0}

In the 5D theory, there are potential linear divergences in the
coupling of the radion to the gauge boson zero modes.  In this
appendix, we show that the linear divergences in fact cancel once the
theory is renormalized.

For the purposes of this demonstration, we work with a simpler theory, a bulk scalar version of QED.
The action is given by
\begin{equation}
S = \int d^5 x - \frac{1}{4} (1+\delta_A (z)) F_{MN}^2 + | D_M \phi
|^2 - M^2 | \phi |^2,
\end{equation}
where we have included the counterterm for divergences in the 5D field
strength one-loop effective action.

The bulk operator which determines the radion coupling to the bulk
scalar field is given by
\begin{equation}
3 T_{55} g^{55} - \Tr T_{MN} = -6 \left( \frac{z}{R} \right)^2 |\nabla_5 \phi |^2 + ( d-4) | D_R \phi
 |^2,
\label{Or}
\end{equation}
where we have set $\Tr g_{\mu \nu}=d$ to work in dimensional
regularization.

The radion couples to the gauge fields directly through the
counterterm necessary to cancel the one-loop divergence in the gauge
boson self energy:
\begin{equation}
3 T_{55} g^{55} - \Tr T_{MN} = ( 1 + \delta_A(z)) \left( \frac{ 6-
  d}{2}\right) \left( \frac{z}{R}\right)^4 A_\mu \left( - g^{\mu\nu} q^2 + q^\mu q^\nu \right) A_\nu
\end{equation}

At one loop, there are two contributions to the radion coupling.
These are the direct coupling to the counterterm above, and the
triangle diagram involving a coupling of the radion to the bulk scalar.

The counterterm is due to the self energy diagram, which is given by
\begin{equation}
\frac{1}{2} g_5^2 \int \frac{d^d p}{(2 \pi)^d} (2p+q)^\mu (2p+q)^\nu
\int_R^{R'} dz dv \left(\frac{R}{z} \right)^3 \left(\frac{R}{v}
\right)^3 G_{|p|} (z,v) G_{|p+q|} (v,z).
\label{eq:counterterm}
\end{equation}
To study the most divergent contribution to the self energy diagram,
we consider the high momentum limit of the full 5D propagator:
\begin{equation}
G_{p} (z,z') \approx \left( \frac{z z'}{R^2} \right)^{3/2} \frac{
  \cosh{(p (z-R))} \cosh{(p (R'-z'))}}{p \sinh{(p (R'-R))}}.
\end{equation}

Performing the integral over $v$, and considering only the leading
term in the $1/|p|$ expansion, Equation~(\ref{eq:counterterm}) reduces to
\begin{equation}
\frac{1}{2} g_5^2 \int \frac{d^d p}{(2 \pi)^d} (2p+q)^\mu (2p+q)^\nu \int_R^{R'} dz \frac{1}{2 |p| (|p|+|q|) (2 |p|+|q|)}.
\label{eq:pexp}\end{equation}
We then expand Eq.~(\ref{eq:pexp}) about small external momentum $q$,
and consider only the terms which are second order in $q$.  The lower
order terms in the expansion lead to cubic divergences which vanish
due to the 5D Ward identity.

After this expansion, and utilizing the symmetries of the momentum
integrals, the self energy diagram becomes
\begin{eqnarray}
&&\frac{1}{24} g_5^2 \left[ -g^{\mu \nu} q^2 \left( \frac{9}{d} -
    \frac{30}{d (d+2)} \right) + q^\mu q^\nu \left( 3 - \frac{18}{d} +
    \frac{60}{d (d+2)} \right) \right] \int_R^{R'} dz \int \frac{d^d
    p}{(2 \pi)^d} \frac{1}{|p|^3} \arline
&&\equiv \frac{1}{24} g_5^2 P^{\mu\nu}_{\rm SE} (d) \int_R^{R'} dz
    \int \frac{d^d p}{(2 \pi)^d} \frac{1}{|p|^3},
\end{eqnarray}
where $P^{\mu\nu}_{\rm SE} = (- g^{\mu\nu} q^2 + q^\mu q^\nu)$ both
when $d=3$ and $d=4$.  The counterterm contribution to the radion
operator is then given by
\begin{equation}
-\frac{6-d}{24} \left( \frac{z}{R} \right)^5 g_5^2 A_\mu
 P^{\mu\nu}_{\rm SE}(d) A_\nu \int \frac{d^d p}{(2 \pi)^d}
 \frac{1}{|p|^3}.
\end{equation}
We now turn our attention to the triangle diagram where the radion
couples to gauge bosons through a scalar loop.  The one loop matrix
element of the scalar contribution to the radion operator between two
gauge fields is given by
\begin{equation}
6 g_5^2 \left( \frac{z}{R} \right)^2  \int \frac{d^d p}{(2 \pi)^d}
(2p+q)^\mu (2p+q)^\nu \int_R^{R'} dw dv \left(\frac{R}{v} \right)^3
\left(\frac{R}{w} \right)^3 \pd_z G_{|p|} (v,z) \pd_z G_{|p|} (z,w)
G_{|p+q|} (v,w). \label{triangle2}\end{equation} Note that the
last term in Eq.~(\ref{Or}) does not contribute after imposing the
equation of motion for $\phi$.

Performing the integrals along the extra dimension, and expanding in
small external momentum $q$, Eq.~(\ref{triangle2}) becomes
\begin{equation}
\frac{1}{8} g_5^2 \left( \frac{z}{R} \right)^5  \left[
-g^{\mu \nu} q^2 \left( \frac{6}{d} - \frac{15}{d (d+2)} \right) +
q^\mu q^\nu \left( 3 - \frac{12}{d} + \frac{30}{d (d +2) } \right)
\right]
\int \frac{d^d p}{(2 \pi)^d} \frac{1}{|p|^3}.
\end{equation}
Thus the triangle diagram contributes the following to the radion
operator at one loop:
\begin{equation}
\frac{1}{8} g_5^2 \left( \frac{z}{R} \right)^5 A_\mu
P^{\mu\nu}_{\bigtriangleup} (d) A_\nu \int \frac{d^d p}{(2 \pi)^d}
\frac{1}{|p|^3},
\end{equation}
where $P^{\mu\nu}_{\bigtriangleup} = (- g^{\mu\nu} q^2 + q^\mu q^\nu)$ both when $d=1$ and $d=3$.

Summing up the two contributions, we find
\begin{eqnarray}
&&\frac{1}{8} g_5^2 \left( \frac{z}{R} \right)^5 A_\mu \left[ P^{\mu\nu}_{\bigtriangleup} (d) - \frac{6-d}{3} P^{\mu\nu}_{\rm SE} \right] \int \frac{d^d p}{(2 \pi)^d} \frac{1}{|p|^3} \arline
&&=- \frac{1}{4} g_5^2 \left( \frac{z}{R} \right)^5 A_\mu \left[ -
    g^{\mu\nu} \frac{3 d-7}{d (d+2)} + q^\mu q^\nu \frac{d^2-4d+14}{d
      (d+2)} \right] \left(\frac{3}{2} - \frac{d}{2}\right) \int
  \frac{d^d p}{(2 \pi)^d} \frac{1}{|p|^3}\arline
&&\
\end{eqnarray}
The momentum integral is proportional to $\Gamma(3/2 - d/2)$,
signalling a linear divergence, however, the coefficient is
proportional to $(3/2 - d/2)$.  Combining this with the
$\Gamma$-function gives a result which is convergent even as
$d \rightarrow 4$.  Thus this contribution is in fact finite.  Note
however that this expression is not transverse as $d \rightarrow 4$.  This is
because an expansion to higher order in the external momentum $q$ is
necessary to capture the complete finite contribution.

\end{document}